\documentclass[final,english]{bullsrsl}[2022/06/15]

\usepackage{aadefs}
\usepackage[latin1]{inputenc}
\usepackage[T1]{fontenc}

\usepackage{natbib} 
\usepackage{graphicx}

\begin{document}
\title{Optical polarisation study of Galactic Open clusters}

\author[affil={1,2}, corresponding]{Namita}{Uppal}
\author[affil={1}]{Shashikiran}{Ganesh}
\author[affil={3}]{Santosh}{Joshi}
\author[affil={3}]{Mrinmoy}{Sarkar}
\author[affil={1}]{Prachi}{Prajapati}
\author[affil={3}]{Athul}{Dileep}
\affiliation[1]{Astronomy and Astrophysics division, Physical Research Laboratory, Ahmedabad-380009, India}
\affiliation[2]{Department of Physics, Indian Institute of Technology, Gandhinagar-382355, India}
\affiliation[3]{Aryabhatta Research Institute of Observational Sciences, Manora Peak, Nainital-263001, India}
\correspondance{namita@prl.res.in, namita.uppal@iitgn.ac.in}
\maketitle


%

\begin{abstract}
Dust is a ubiquitous component in our Galaxy. It accounts for only 1\% mass of the ISM but still is an essential part of the Galaxy. It affects our view of the Galaxy by obscuring the starlight at shorter wavelengths and re-emitting in longer wavelengths. Studying the dust distribution in the Galaxy at longer wavelengths may cause discrepancies due to distance ambiguity caused by unknown Galactic potential. However, another aspect of dust, i.e., the polarisation of the background starlight, when combined with distance information, will help to give direct observational evidence of the number of dust clouds encountered in the line of sight. We observed 15 open clusters distributed at increasing distances in three lines of sight using two Indian National facilities. The measured polarisations results used  
to scrutinize the dust distribution and orientation of the local plane of sky magnetic fields towards 
selected
directions. The analysis of the stars observed towards the distant cluster King 8 cluster shows two foreground layers at a distance of $\sim 500$ pc and $\sim$ 3500 pc. Similar analysis towards different clusters also results in multiple dust layers.
\end{abstract}

\section{Introduction}
Dust plays a significant role in the Galaxy, influencing various physical and chemical processes.  It is predominantly concentrated within the disc of edge-on galaxies like the Milky Way, particularly within the spiral arms. As a result, dust serves as a valuable tracer for mapping the Galactic disc, specifically the spiral arms. \citet{Drimmel2001} presented a three-dimensional Galactic dust distribution model using COBE/DIRBE data. However, the investigation of dust distribution in the Galaxy using longer wavelengths depends on the kinematic distances of dust clouds, which can introduce significant uncertainties \citep{Sanna2009}. These uncertainties may potentially lead to erroneous interpretations of the Galactic structure. Therefore, rather than directly tracing dust distribution in longer wavelengths, an alternative approach is to map the structure using the indirect properties of dust, such as extinction and polarisation. 
The availability of multi-wavelength all-sky surveys has enabled numerous efforts to construct three-dimensional extinction maps of the Galaxy \citep[e.g.,][]{Sale2014, lallement2018, Rezaei2018, Green2019}. However, it is important to note that these extinction maps are model-based and rely on certain assumptions, which may introduce inherent biases.

The dust which is responsible for the extinction may also cause polarisation.  The phenomenon of interstellar (ISM) polarisation was first observed independently by \citet{Hall1949} and \citet{Hiltner1949}. The ISM polarisation arises when starlight from an unpolarised star passes through optically anisotropic dust grains that are partially aligned with a magnetic field \citep{Davis1949}
The simplest anisotropy could arise from the asymmetry in the grain size, such as elongated cylindrical type grains. Thus the starlight polarisation results from differential extinction by the asymmetric, elongated grain, partially aligned with a magnetic field. 
Similar to extinction, polarisation, in combination with distance information, can also be utilised to infer the dust distribution and investigate the underlying structure. This idea has been utilised in various studies at small scales to find the dust distribution in different lines of sight \citep[e.g.,][]{Eswaraiah2012, Uppal2022, Pelgrims2023}. Unlike extinction, which is a derived quantity, polarisation results from direct observations and provides information on dust distribution without relying on any model parameters. While a significant amount of effort has been devoted to comprehending the structure of the Milky Way galaxy through the use of extinction maps, there are only limited dedicated surveys that provide information on interstellar polarisation \citep{Heiles2000, Planck2020, Clemens2020DR4, Versteeg2023}. In addition, these surveys cover only a limited portion of the sky.
To trace the Galactic structure in greater detail, it is crucial to conduct more polarisation observations in the Northern Hemisphere. By systematically observing polarisation along different lines of sight, we can explore both small- and large-scale structures within the disc of our galaxy.

Open clusters are ideal candidates to carry out polarisation observations because all the member stars have common properties such as distance, age, and proper motion. Moreover, a bunch of stars present at similar distances and locations in the sky will provide statistically stronger results as compared to randomly selected stars in the sky. Out of the 7200 sample \citep{Hunt2023} of open clusters detected from Gaia DR3 \citep{GaiaDR3}, only ~40 clusters have been observed polarimetrically to date. We require polarisation observations of a large number of open clusters to make 3D dust or magnetic field tomography. In this paper, we highlight the polarisation observation of 11 open clusters and their role in mapping the dust distribution.

\section{Observations}
The polarisation observations of Galactic open clusters in the Northern sky were conducted over a period ranging from January 2021 to October 2022. These observations were carried out utilising two prominent national observation facilities: the 1.2 m telescope at Mount Abu Observatory, operated by the Physical Research Laboratory (PRL), and the 1.04 m Sampurnanand telescope at the Aryabhatta Research Institute of Observational Sciences (ARIES), located in Nainital.
\begin{table}[!h]
	\centering
\begin{minipage}{150mm}
	\caption{Observation detail of the targeted open clusters.}\label{tab:1}
 \end{minipage}
\bigskip
	\begin{tabular}{lccccr} 
        \hline
	Cluster & $\ell$ & $b$ &  distance (kpc) & instrument & total exposure (s)\\
		\hline
            Berkeley 47 & 52.545 & -0.041 & 2.3 &AIMPOL & 600 \\
            Kronberger 79 & 54.180 &-0.611 & 5.6 & EMPOL & 1440\\
            Kronberger 69  & 68.516	& 0.433 & 2.6 &AIMPOL & 1500 \\
            Kronberger 54 & 69.106 & 0.523 & 4.4 & EMPOL & 2880 \\
            Berkeley 49 & 70.979 & 2.578 & 3.2 &AIMPOL & 900 \\
            Czernik 3 & 124.256 & -0.058 & 4.5 & EMPOL & 1920 \\
		Kronberger 1 & 173.107 & 0.046 & 2.1 & AIMPOL & 720 \\
		Berkeley 69 & 174.440 & -1.852 & 3.5 & AIMPOL & 1680 \\
		King 8 & 176.384 & 3.101 & 6.2 & AIMPOL & 1500 \\
		Berkeley 71 & 176.630 & 0.894 & 3.9  & AIMPOL & 1200\\
		Berkeley 19 & 176.920 & -3.619 & 6.1 &AIMPOL & 1500 \\
            
		\hline
	\end{tabular}
\end{table}
A comprehensive set of 11 clusters was observed, encompassing three distinct lines of sight. A summary of the observational details, including the Galactic longitude, latitude, distance \citep{gaiaDR2OC}, the instrument used for observation, and total exposure time for each cluster is provided in Table \ref{tab:1}. We targeted clusters in a similar line of sight but distributed according to distance so that we could get continuous signatures of the number of dust clouds in a particular line of sight. For the observations, we have used an EMCCD-based POLarimeter (EMPOL) as a back-end instrument on a 1.2 m PRL telescope and an ARIES Imaging POLarimeter (AIMPOL) on 1.04 m Sampuranand telescope.  Detailed information regarding these instruments and the observations conducted with them is presented in the following subsections.

\subsection{EMPOL}
EMPOL is a single-channel polarimeter employing a rotating half-waveplate (HWP) to modulate the star intensity for polarisation measurement and a wire grid polariser, which acts as an analyser. The rotating HWP is driven by a stepper motor, allowing it to complete its rotation in 48 steps. Each step of the HWP corresponds to a rotation of 7.5 degrees and is accompanied by 0.5-second exposures. A fixed HWP is also used just below the rotating HWP to compensate for the wavelength dependence of the position angle of the HWP. The instrument is equipped with a twelve-spaced filter wheel housing Johnson-Cousins broadband filters (B, V, R, and I), as well as Sloan filters (g, r, i, z) and three narrow-band filters. Finally, the light is detected at the EMCCD, which is used for its fast readout capability with minimal read noise and on-chip gain functionality. The advantage of such rapid readout is the ability to measure the flux at multiple angles, reducing systematic errors in the resulting polarisation measurements.
\begin{figure}[!h]
\centering
\includegraphics[scale = 0.25]{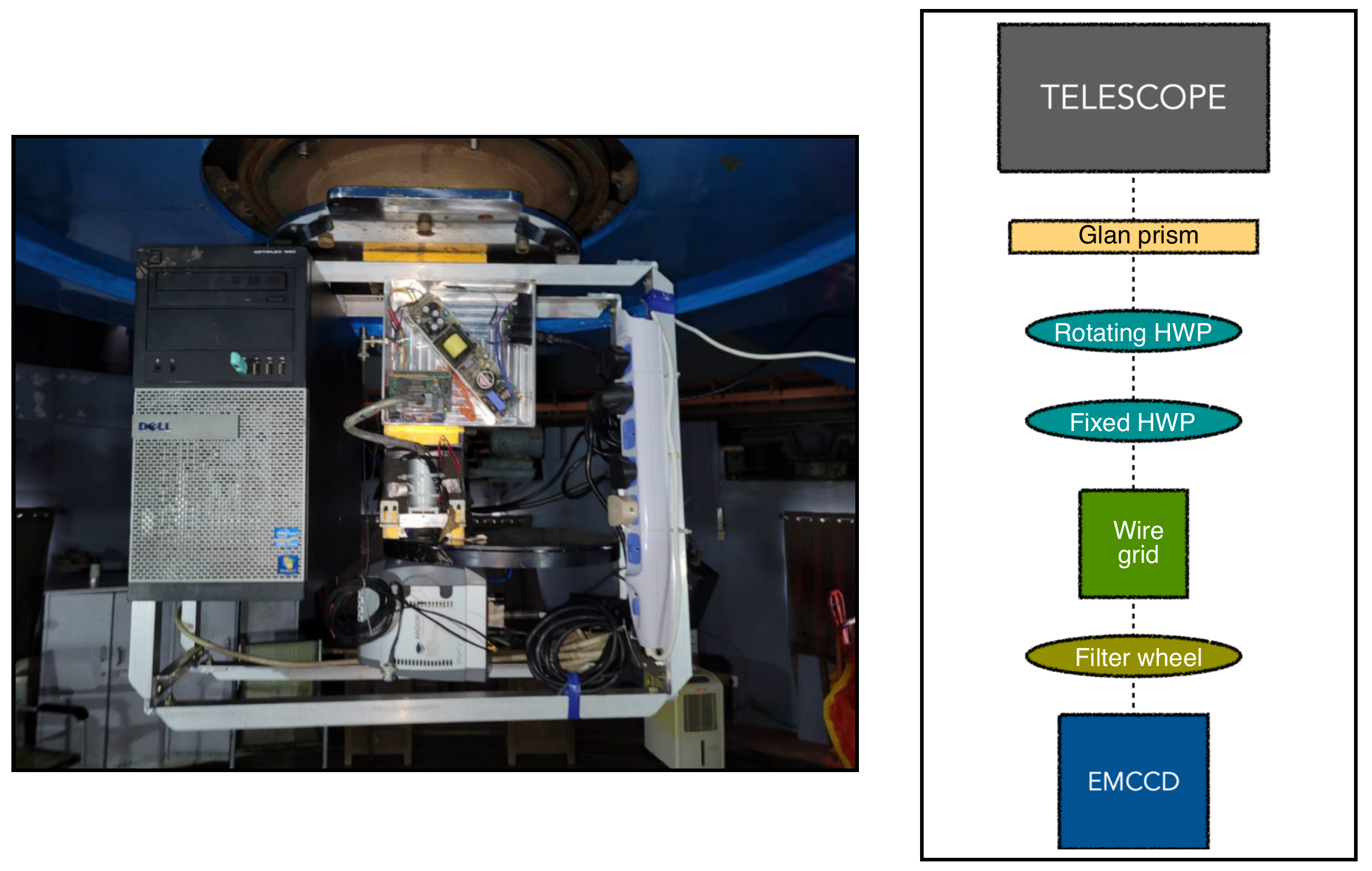}
\bigskip
\begin{minipage}{12cm}
\caption{EMPOL instrument mounted on 1.2 m Mount Abu telescope in the left panel and the right panel represents the schematic optical layout of the instrument. }\label{fig:3}
\end{minipage}
\end{figure}
However, this setup does have a limitation: it requires at least one bright star to be present in the field within the 0.5-second exposure, which can be utilised for stacking to improve the signal-to-noise ratio.  The instrument provides a small field of view of $\sim 3^\prime \times 3^\prime$. Thus the core region of the clusters is observed in a single pointing. Further details of the instrument can be found in \citet{EMPOL} and \citet{Uppal2022}. Figure \ref{fig:3} illustrates the EMPOL instrument mounted on the 1.2 m Mount Abu telescope (on the left) and its schematic optical layout (on the right panel).

Throughout the observation season of 2021-2022, we conducted regular observations with EMPOL, utilising 3-4 nights per month. These observations focused on several open clusters within our target range. Among them, we successfully observed the core region of the distant cluster; Czernik 3 (located in the second Galactic quadrant) using the Sloan i-band and two clusters (Kronberger 79, and Berkeley 54) in the first Galactic quadrant using the R-band.  The exposure time at each position of HWP is limited to 0.5 sec. So, to achieve an effective exposure of, say, 10s, a total of 1008 images were recorded. Thus the total exposure would be $10 s\times 48 = 480 s$. The total exposure time for various clusters observed is listed in Table 1. Furthermore, we employed $4 \times 4$ on-chip binning and an electron multiplicative gain (EMGAIN) of 20 to obtain sufficient counts for faint stars.

\subsection{AIMPOL}
AIMPOL is a dual-channel polarimeter that utilises a Wollaston prism as an analyser, in contrast to the wire grid used in the EMPOL. The Wollaston prism splits the incoming light into ordinary (o-ray) and extra-ordinary (e-ray) components.  These two rays fall on the same CCD plane separated by $\sim 34$ pixels. The instrument encompasses an achromatic rotating HWP, that rotates in 4 steps, i.e., at $0^\circ$, $22.5^\circ$, $45^\circ$, and $67^\circ.5$  to modulate the intensity of the star.
\begin{figure}[!h]
\centering
\includegraphics[scale = 0.25]{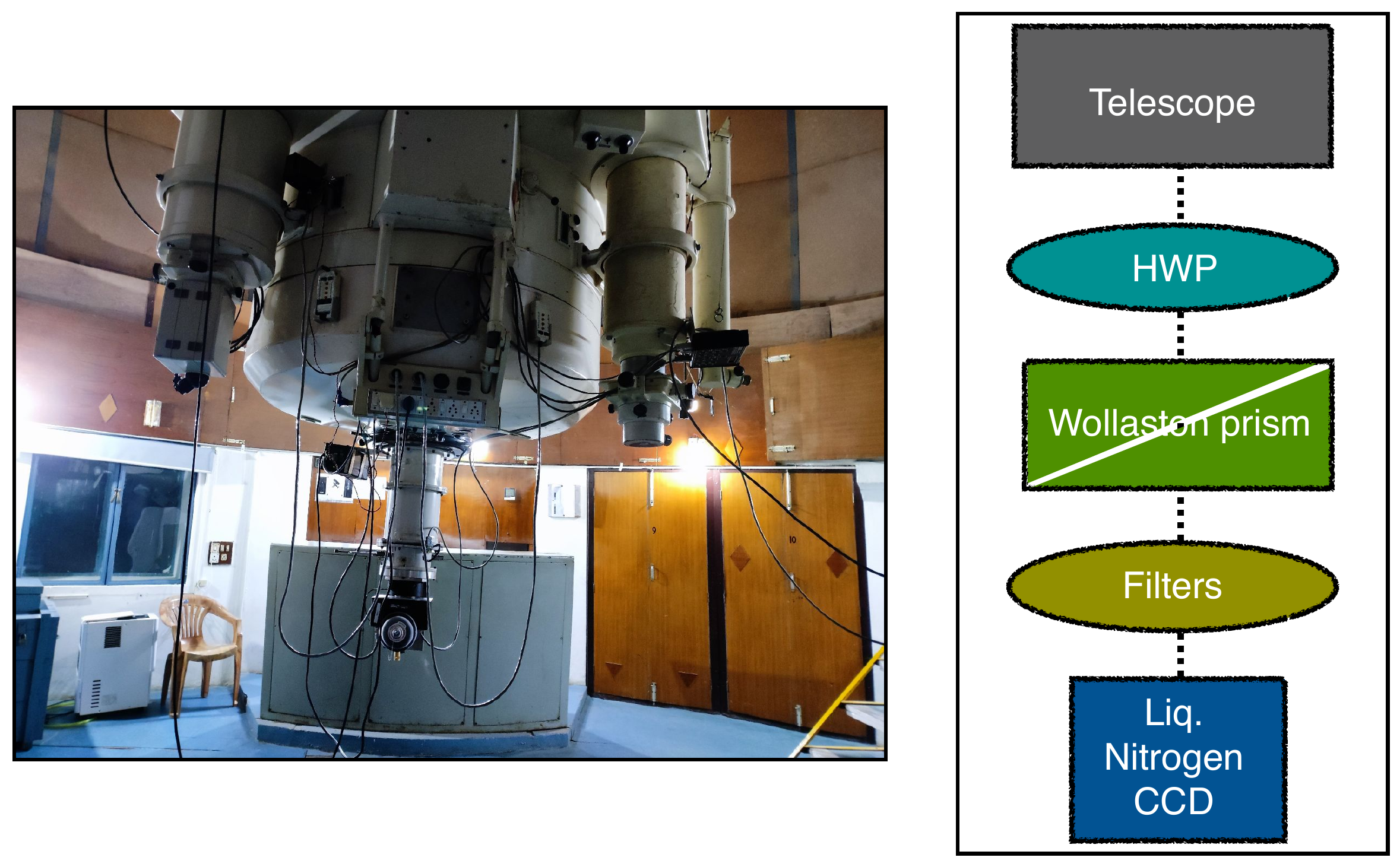}
\bigskip
\begin{minipage}{12cm}
\caption{AIMPOL instrument mounted on 1.04 m Sampurnanand telescope in the left panel and the right panel represents the schematic optical layout of the instrument.}\label{fig:4}
\end{minipage}
\end{figure}
A liquid nitrogen-cooled 1340 $\times$ 1300 pylon CCD is used to detect the modulated light coming from the celestial object giving a larger field view of $\sim 8^\prime$ in diameter. In addition, the CCD offers various readout speeds (50, 100, 200, 500 kHz, 1, 2, 4 MHz) and gain options (low, medium, and high). Further details of the instrument can be found in \citet{Rautela2004} and \citet{Pandey2023}. A view of the AIMPOL instrument mounted on a 1.04 m Sampurnanand telescope and its schematic optical layout is presented in Figure \ref{fig:4}.

We submitted proposals to observe several open clusters, primarily focusing on nearby clusters with larger sizes that could be fully covered within single pointing of AIMPOL. In total, we were allocated 11 nights spread over two observation cycles. We conducted observations of 8 Galactic open clusters using AIMPOL in the R-band, with a readout speed of 100 kHz and the medium gain option. Among the observed clusters, five are located in the anticenter direction: Kronberger 1, Berkeley 69, King 8, Berkeley 71, and Berkeley 19. The remaining three clusters (Berkeley 47, Berkeley 49, and Kronberger) are situated in the first Galactic quadrant. The total integration times used for the various clusters are listed in the corresponding column of Table \ref{tab:1}.

\section{Data Reduction}
\begin{figure}[!h]
\centering
\includegraphics[scale = 0.23]{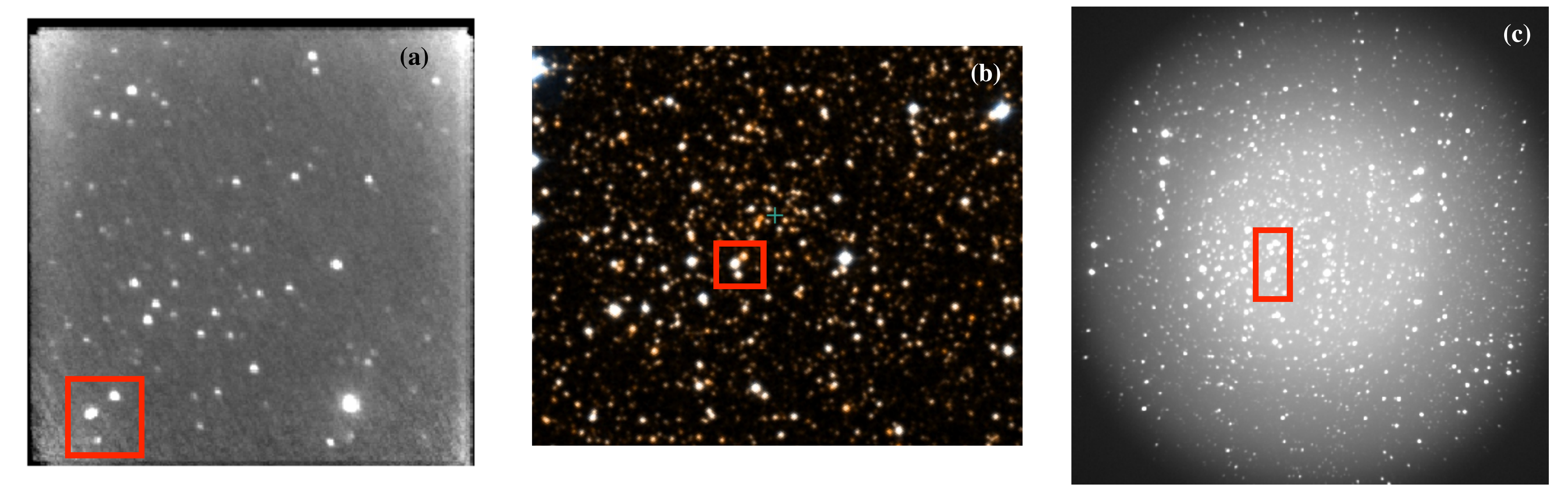}
\bigskip
\begin{minipage}{12cm}
\caption{Comparison of the images taken from EMPOL (panel a) and AIMPOL (panel c) with the DSS image (panel b) of the Berkeley 69 cluster. The red box in each panel highlights the common stars in three panels. }\label{fig:5}
\end{minipage}
\end{figure}

\begin{figure}
\centering
\includegraphics[scale = 0.18]{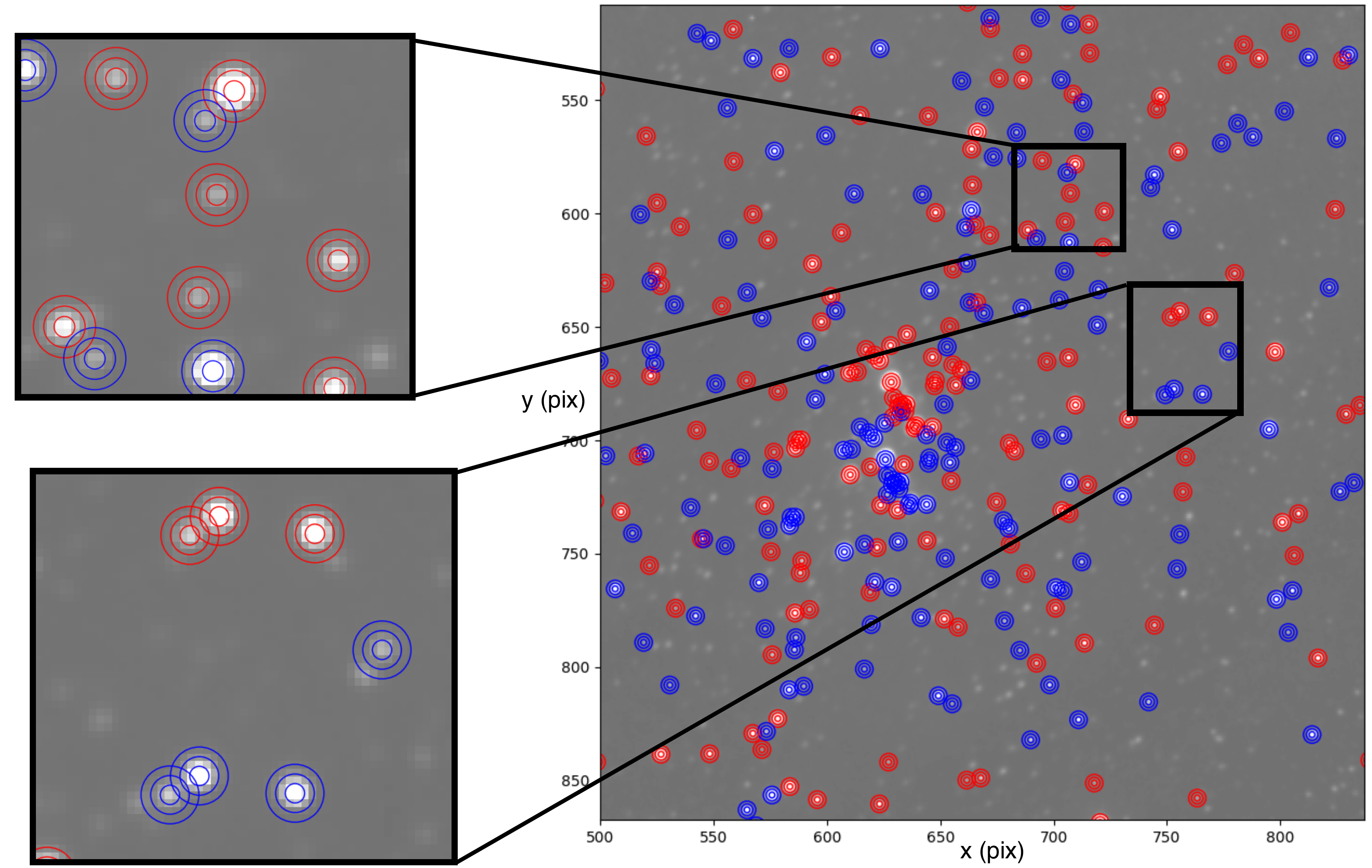}
\bigskip
\begin{minipage}{12cm}
\caption{An image showing the aperture-overlapping issue in AIMPOL data. The concentric circles represents the chosen aperture according to 1, 2, and 3 $\times$ FWHM, in red for o-ray and blue for e-ray. }\label{fig:6}
\end{minipage}
\end{figure}

The observed data were reduced and analysed using self-scripted Python routines. The basic data reduction tasks include - bias subtraction, shifting and stacking of images to increase the signal-to-noise ratio.  For EMPOL, shifting and stacking were performed in cycles of 48 frames, while for AIMPOL, the cycles consisted of 4 frames to obtain a final stack corresponding to each angle of the HWP. The image coordinates of stars in the field were obtained using a web-based `astrometry.net' service. In cases where the service failed to produce results, we employed the \textit{SExtractor} software \citep{sextractor} for astrometry. The astrometry is followed by the photometry of all the objects in the field at each HWP position using aperture photometry with \textit{photutils} package of \citep{photutils} of \textit{Astropy} \citep{astropyII}. AIMPOL being a dual channel polarimetry, the e-ray and o-ray sources are focused on the same CCD plane, giving two images of a single object as shown in panel (c) of Figure \ref{fig:5}.
It is crucial to separate the e-ray and o-ray images of the stars. Since the separation of o-ray and e-ray images on the CCD plane is fixed, it is possible to separate all the e-ray sources from the o-ray sources by identifying a single e-ray and o-ray pair in the field. Due to the crowded nature of open cluster fields, there is a finite possibility of intensity overlap between close stars. The situation becomes more complicated in AIMPOL, where the e-ray and o-ray images of different stars are present on the same CCD plane. Consequently, the o-ray of one star may also overlap with the e-ray of another star and vice-versa, adding to the complexity of the analysis (see Figure \ref{fig:6}). To overcome this issue multiple apertures were defined between $1\times$ FWHM to $3\times$ FWHM. For EMPOL, which is a single-channel polarimeter, the modulation affects the intensity of the incoming light, and the polarisation is measured based on the intensity of the star at each HWP position relative to the maximum intensity of the object. We applied aperture corrections obtained from isolated stars to mitigate the effects of overlapping in this case. In contrast, for AIMPOL, the dual-channel polarimeter, the relative intensities of the two images at different HWP angles were required to measure polarisation.

We calculated the normalized Stokes parameters, q (Q/I) and u (U/I), by fitting Equations \ref{eq:1} and \ref{eq:2} to the intensity ($I_j^\prime$) or modulation factor ($R_\alpha$) of the object at each HWP position or angle for EMPOL and AIMPOL, respectively. Here, j represented the number of rotation steps of the HWP, ranging from 0 to 47 for EMPOL and from 0 to 3 for AIMPOL.
\begin{equation}\label{eq:1}
    I_j^\prime = \frac{1}{2} [I\pm Q\cos4\alpha_j \pm U\sin4\alpha_j]
\end{equation}
\begin{equation}\label{eq:2}
    R(\alpha_j) =  \frac{\frac{I_e}{I_o} \times F-1}{\frac{I_e}{I_o}\times F+1} = P_{obs} \cos (2\theta - 4 \alpha_j) = q \cos 4\alpha_j  + u \sin 4\alpha_j 
\end{equation}
Here, the factor F is based on the fraction of o-ray and e-ray intensities, i.e., $\frac{I_o}{I_e}$, which may be related to the difference in the response of the system to e-ray and o-ray and/or the different response of CCD as a function of position on the surface \citep{IMPOL}. This factor is estimated as follows.
    \begin{equation}
        f = \left[\frac{I_o(0^\circ)}{I_e(45^\circ)} \times \frac{I_o(45^\circ)}{I_e(0^\circ)}\times \frac{I_o(22.5^\circ)}{I_e(67.5^\circ)}\times\frac{I_o(67.5^\circ)}{I_e(22.5^\circ)}\right]^{\frac{1}{4}}
    \end{equation}
    
The linear degree of polarisation ($P_{obs}$) and polarisation angle ($\theta$) are calculated from measured q and u parameters from equation \ref{eq:3} and \ref{eq:4}.
\begin{equation}\label{eq:3}
P_{obs} = \sqrt{q^2+u^2}
\end{equation}
\begin{equation}\label{eq:4}
\theta = \frac{1}{2}\tan^{-1}\left(\frac{u}{q}\right)
\end{equation}

The errors in the degree of polarisation and polarisation angle were derived using fundamental error propagation methods on Equations \ref{eq:3} and \ref{eq:4}.
All the described tasks for image reduction are made fully automated using various Python modules. 

\section{Results and discussions}
We measure the polarisation of at least the core region of eleven clusters present along three lines of sight. The polarisation measurements towards each cluster direction are crossmatched with the Gaia EDR3 distance catalogue to get the accurate distance of the corresponding stars. The variation of the degree of polarisation, and polarisation angle as a function of distance is analysed to find a number of foreground dust layers. It is expected that when the light from stars encounters dust layers at various distances in the line of sight, the degree of polarisation of background starlight is expected to exhibit a jump. This variation is caused by the orientation of the magnetic field within each dust cloud. If the magnetic field is uniform in all dust clouds, the degree of polarisation is expected to increase as the starlight passes through each dust layer. On the other hand, if the orientation of the magnetic field shows deviation, then the starlight becomes depolarised. The number of such changes in the slope of the degree of polarisation and extinction as a function of the distances to the stars corresponds to the number of foreground dust layers. Therefore, by analysing polarisation with distance, we can study the dust distribution along the line of sight.

\begin{figure}[!h]
\centering
\includegraphics[scale = 0.3]{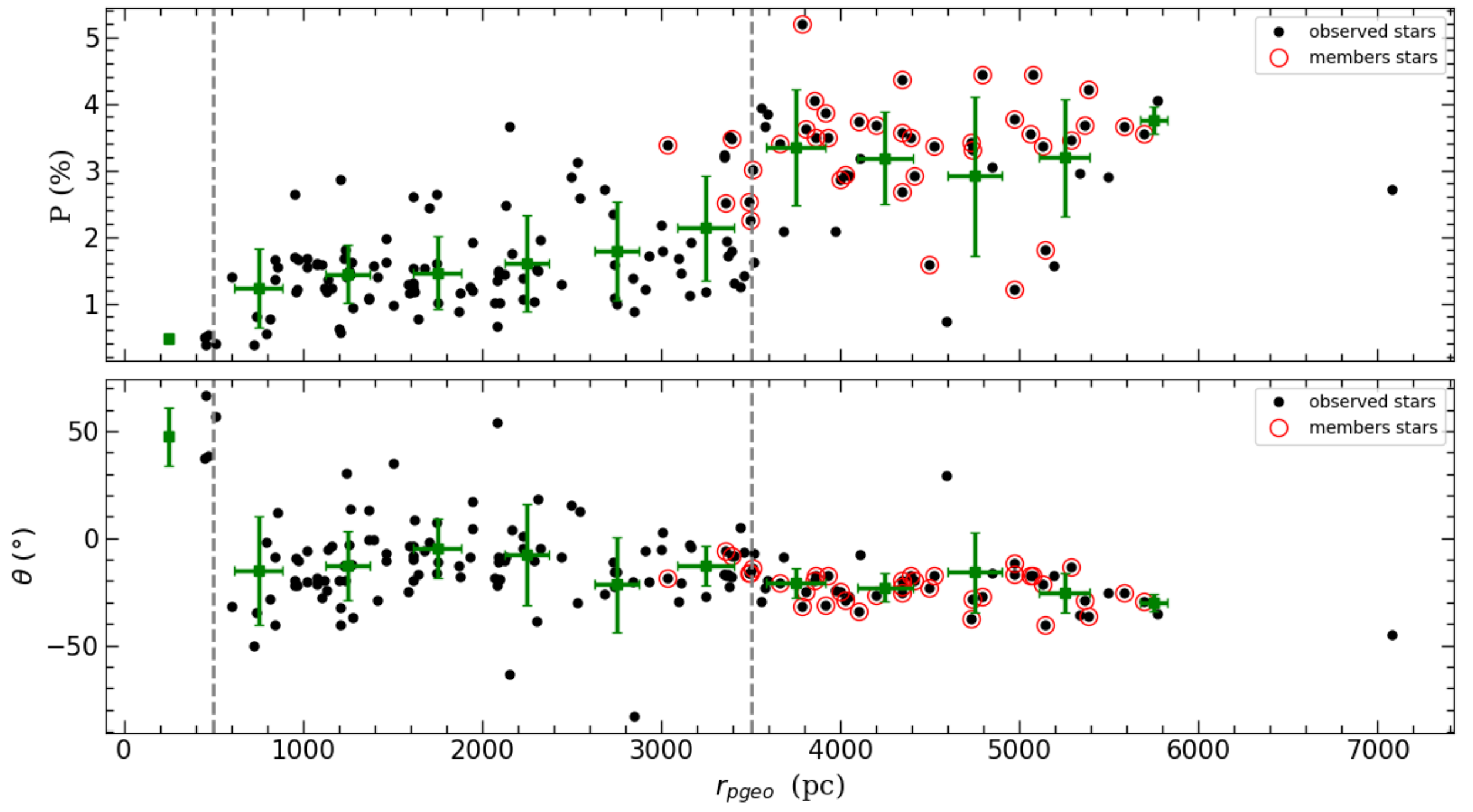}
\bigskip
\begin{minipage}{12cm}
\caption{Variation of degree of polarisation as a function of Gaia EDR3 Bayesian based distance $r_{pgeo}$ \citep{bailer2021}. The vertical grey lines correspond to the distance of predicted dust layers.}\label{fig:7}
\end{minipage}
\end{figure}

The detailed analysis and possible interpretation for one cluster, Czernik 3, was drawn in \citet{Uppal2022}. Similar analyses were performed for each observed cluster, and the polarisation results and interpretations are presented elsewhere. The variation of the degree of polarisation as a function of distance for the King 8 cluster is shown with black points in figure \ref{fig:7} for illustrative purposes. The member stars having a probability of more than 50\% according to the criteria described in \citet{gaiaDR2OC} are depicted in red open circles. In order to clearly identify the number of dust layers, the average degree of polarisation and polarisation angle is determined in 500 pc bins of distance and is shown as green squares with error bars representing dispersion in the measured quantity. The degree of polarisation shows an abrupt increase at a distance of $\sim$ 500 pc and $\sim$ 3500 pc (see grey vertical lines). We also observed that the dispersion in polarisation angle reduces at respective distances. Thus jump in the degree of polarisation and polarisation angle depicts the existence of two foreground dust layers. 

Figure 11 of \citet{Uppal2022} highlights the detection of the inter-arm region between the Local arm and Perseus arm from polarisation. So, the abrupt changes in the polarisation towards different lines of sight may give signatures of spiral arms, at least towards the low extinction directions. The high extinction field may represent the patchy dust distribution. The large-scale dust distribution required continuous observations of various clustered regions covering the Galactic plane. 
In future, we hope to cover more areas of the Galactic plane to study large-scale dust distribution.

\begin{acknowledgments}
NU acknowledges support from BINA-2 for participating in the meeting. We thank the anonymous referee for reviewing our manuscript and their suggestions. We also thank the time allocation committee of the 1.2 m Mount Abu telescope and the 1.04 m Sampurananand telescope for the allotted time. We are grateful to the dedicated staff at both observatories for their valuable assistance during the observations. Work at Physical Research Laboratory is supported by the Department of Space, Govt. of India. We Specially thank Mr Arvind K. Dattatrey for his continuous support throughout the observing epochs with AIMPOL.  We would also like to acknowledge the assistance of Dr Gaurav Singh, Indian Institute of Astrophysics, Bangalore, and Ms Gulafsha Choudhury, Assam University, Silchar, for useful discussion regarding AIMPOL observations.  A part of this work has used data from the European Space Agency (ESA) mission Gaia, processed by the Gaia Data Processing and Analysis Consortium (DPAC). Funding for the DPAC is provided by national institutions, particularly the institutions participating in the Gaia Multilateral Agreement. This research used Astropy, a community-developed core Python package for Astronomy. We have also used the VizieR catalogue access tool, CDS, Strasbourg, France.
\end{acknowledgments}
\begin{furtherinformation}
\begin{orcids}
\orcid{0000-0001-5814-4558}{Namita}{Uppal}
\orcid{0000-0002-7721-3827}{Shashikiran}{Ganesh}


\end{orcids}

\begin{authorcontributions}
NU and SG collaborated on generating the underlying idea of the manuscript. NU performed the observations with valuable assistance from PP and AD.  NU and SG have performed the data reduction, analysed and interpreted the results. NU and SG were also involved in writing the manuscript, with all authors contributing to its refinement.
MS served as the Principal Investigator (PI) of the AIMPOL observation proposal, with NU, SG, and SJ as Co-Investigators (Co-Is). All authors have reviewed and approved the final version of the manuscript.
\end{authorcontributions}

\begin{conflictsofinterest}
The authors declare no conflict of interest.
\end{conflictsofinterest}

\end{furtherinformation}

\bibliographystyle{bullsrsl-en}

\bibliography{extra}

\end{document}